\documentstyle[aps,preprint]{revtex}
\begin{document}
\title{Properties of a continuous-random-network model
for amorphous systems}
\author{Yuhai Tu, J. Tersoff, G. Grinstein}
\address{IBM Research Division, T. J. Watson Research Center,
P. O. Box 218, Yorktown Heights, NY 10598}
\author{David Vanderbilt}
\address{Department of Physics, Rutgers University, P. O. Box 849,
Piscataway NJ 08855}

\date{\today}

\maketitle

\begin{abstract}
We use a Monte Carlo bond-switching method to study systematically
the thermodynamic properties of a ``continuous random network" model,
the canonical model for such amorphous systems as a-Si and a-SiO$_2$.
Simulations show first-order ``melting" into an amorphous state,
and clear evidence for a glass transition in the supercooled liquid.
The random-network model is also extended to study heterogeneous
structures, such as the interface
between amorphous and crystalline Si.
\end{abstract}


Amorphous materials have been intensely studied
because of both their technological importance
and fundamental interest.
Due to the complexity of these systems,
theoretical studies have relied mainly on numerical simulations.
The standard method is molecular dynamics (MD)
\cite{MDepot,MDabinitio,lSiO2,MDoxide},
where atoms are represented by point particles
with more or less realistic interactions,
and the equations of motion are integrated numerically.

Most studies of glassy behavior in amorphous systems have involved
hard-sphere or Lennard-Jones models for metallic glasses.
However, the most important and ubiquitous amorphous materials
are network glasses such as a-SiO$_2$.
These may also be studied by MD,
but large-scale or long-time simulations are limited to the few systems
where suitable classical models are available for the atomic interactions
\cite{MDepot,MDoxide}.
Moreover, MD methods are inherently less efficient for
strong network-forming materials,
because there is a large energy barrier to breaking
and reforming bonds in these systems.
Most of the computer time is therefore spent in following local
vibrations, waiting for the infrequent bond-switching events.

An alternative is to use the canonical model for such network glasses:
a continuous random network (CRN) of atoms and bonds \cite{Zach}.
The most realistic available models of amorphous Si are of this type.
They have been generated by
Wooten, Winer and Weaire \cite{W3} (hereafter WWW),
using an ingenious Monte Carlo (MC) approach
for generating random networks by bond switching.

Here we show that this approach can be extended
to directly study the thermodynamic properties of
disordered materials.
The system is represented explicitly as a CRN, i.e., as a
set of atoms and a ``neighbor list" specifying
which pairs of atoms are connected by bonds.
Equilibrium or quasi-equilibrium properties are determined
by MC sampling, using the bond-switching move
introduced by WWW.

The specific model system studied here has four bonds to each atom,
as for a-Si.
But we view it principally as a generic CNR model,
and we systematically study its phase diagram and glass transition.
This approach, and the phase structure we find,
should apply also to more complicated
covalent amorphous solids such as a-SiO$_2$.  In addition, we use
the model to explore the properties of the interface between the
crystalline and amorphous phases of silicon.

The energy of the CRN is given by
a Keating-like valence force model \cite{Keat},
and depends on both
the positions $\{\vec{r}_i\}$ of the atoms and
the set $\psi$ of bonds connecting pairs of atoms:
\begin{eqnarray}
E_{\rm tot}(\psi ,\{\vec{r}_i\}) &=&\sum_{i,j\epsilon
\psi }\frac{1}{2}k_{\theta}b_{0}^{2}(\cos \theta _{ij}
-\cos \theta _{0})^{2}  \nonumber \\
&&+\sum_{j\epsilon \psi }\frac{1}{2}k_{b}(b_{j}-b_{0})^{2}~.
\end{eqnarray}
Here $j$ represents the $j$th bond, $b_j$ is its length,
$\theta_{ij}$ is the
angle between bonds $i$ and $j$ connected to a common atom,
$b_0$ is the preferred bond length,
$\theta_0$ is the preferred bond angle,
and $k_{\theta}$ and $k_b$ are ``spring constants."
In order to focus on the role of network structure,
we write the energy as a function solely of bond topology,
minimizing $E_{\rm tot}$ with respect to
the geometrical coordinates $\{\vec{r}_i\}$:
\begin{equation}
E(\psi)=\min_{\{\vec{r}_i\}} E_{\rm tot}(\psi,\{\vec{r}_i\}) ~~.
\end{equation}

Using $E(\psi)$, we
can study the statistical properties of the system through
MC simulation.  We use the WWW
construction for the local MC moves in the $\psi$ space.  That is,
from an initial configuration $\psi_1$, a bond
is chosen randomly (call it BC), and one
more bond connected to each terminus
is also chosen randomly (bonds
BA and CD), the only constraint being that all four
atoms A, B, C and D must be distinct.
The switching move leading to $\psi_2$ is then simply the cutting of bonds
BA and CD and the formation of new bonds AC and BD.
In this way the system samples topologically distinct
configurations without introducing ``dangling bonds"
or changing the number of bonds to any atom.

For a given temperature $T$,
one uses the Metropolis MC algorithm
to decide whether a given switching move is accepted or
rejected: the move is accepted with
probability $P=\min[1,\exp(-\Delta E/k_{B}T)]$, where $\Delta
E=E (\psi_2)-E (\psi_1)$ is the energy change.
Note that the temperature $T$ applies only to the bond
topology; the system is always in its ground state
with respect to phonons.
This makes the model less realistic in describing
certain aspects of specific systems,
but more generic in distilling the role of network topology
and excluding other issues.

We choose parameter values appropriate for a-Si:
$k_{\theta}=0.647$ eV/\AA$^2$, $k_{b}=9.08$ eV/\AA$^2$,
$b_0=2.35$ \AA, and $\theta_0=109^\circ$.
In addition, it is important that the neighbor list $\psi$
remain consistent with the geometry $\{\vec{r}_i\}$,
with close atoms included in the neighbor list,
and distant atoms excluded.
To guarantee this, we include an extra energy term $E^{\prime}$
in $E_{\rm tot}$ to prevent structures with
``false" neighbors from occurring:
$E^{\prime} =\gamma\sum_{mn} (d_2-|\vec{r}_m-\vec{r}_n|)^3$.
Here $m$ and $n$ label atoms which are neither 1st nor
2nd neighbors in $\psi$, but for which $|\vec{r}_m -\vec{r}_n|$
is actually less than the distance $d_2$=3.84 \AA\ between
next-nearest neighbors in crystalline silicon.
We use $\gamma=0.5$ eV/\AA$^3$.

We begin our simulations in the ground state
of the system, a diamond-structure crystal,
where each atom has four bonds.
At low temperature this crystalline phase is stable,
while at high temperature the crystal ``melts" into
a disordered ``liquid" phase.
Note that this disordered system is still a network,
with the same bond coordination as the crystal.
Good glass-formers such as SiO$_2$ typically share
this characteristic
that the liquid retains the network structure \cite{lSiO2}.

It is difficult to determine the melting point accurately
in a homogenous system,
because of the energy barrier to nucleation of a new phase.
We therefore create a system with a solid-liquid interface,
and study the interface motion as a function of temperature.
For temperatures $T<T_m$, the
crystal phase will invade the amorphous phase, while for $T>T_m$
the reverse occurs.
The transition temperature $T_m$ corresponds to the temperature
where the interface motion vanishes.

The sample is prepared by taking a crystal
and allowing WWW switching in half the cell,
while forbidding switching in the other half.
At sufficiently high $T$, the first half becomes liquid.
To determine the interface motion,
the system is then cooled to the temperature of interest,
and switching is allowed throughout the sample.

We have simulated interface motion in this manner at different
temperatures for systems with $N$=432 atoms, with interfaces oriented along
the [111] direction.
Figure 1 shows snapshots of the atomic positions projected
onto the $(01\bar{1})$ plane, at six successive times,
for each of the two temperatures $k_B T_{1}$=0.50 eV [Fig.\ 1(a)] and
$k_B T_{2}$=0.55 eV [Fig.\ 1(b)].
Fig.\ 1 makes clear that the amorphous central region shrinks at
temperature
$T_{1}$ by recrystallization,
while at $T_2$ the more stable amorphous phase grows
at the expense of the crystal.
We conclude that $T_1 < T_m < T_2$.
From further calculations of this type, we estimate that
$k_B T_m\sim 0.53 \pm 0.01$ eV.

This $T_m$ is a very high temperature, around 6000K.
Note that real Si does not form such a network liquid,
but rather melts into a very different high-coordination
liquid phase at much lower temperature, around 1700K.
As a result, Si is a poor glass-former, and a-Si can only be
formed by processes which are very far from equilibrium.

We have also studied the properties of the homogeneous
disordered network (the ``liquid" phase)
for temperatures both above and below $T_m$, for system size $N$=216.
In particular, the average energy per atom, $E_a (T)$,
of the amorphous phase is determined by
equilibrating for $\sim 2 \times 10^5$ MC steps and then
averaging over $\sim 10^6$ MC steps, for each $T$. To further reduce 
inaccuracy due to statistical fluctuation, 
the result is averaged over typically 10-20 
runs with different random number seeds.
From standard thermodynamic relations, we can then easily calculate the
entropy, $S_a (T)$, and free energy, $F_a (T)$, per atom:
$S_a(T)= S_a (T_0) + \int_{T_0}^T
(1/T)(\partial E_a/\partial T) dT $,
and $ F_a(T) =E_a(T)-TS_a(T) $.
It is particularly
convenient to choose the arbitrary temperature
$T_0$ to be $T_m$,
since $F_a (T_m) = F_c
(T_m)$, where $F_c (T_m)$ is the free energy for the
crystal.
In this model $F_c(T)=0$ to an excellent
approximation for temperatures in the range of interest \cite{FcT}.

The resulting values for $E_a(T)$, $F_a(T)$, and $S_a(T)$
are shown in Fig.\ 2.
When $T<T_m$, $F_a(T)>0$,
so the crystal phase is
thermodynamically preferred, while for $T>T_m$, $F_a(T)<0$,
and the amorphous ``liquid" phase is more stable.
Our simulations clearly indicate that the crystalline phase
is metastable in the region of stability of the amorphous phase, and
vice versa, with a large nucleation barrier separating the phases.

Thus, we are able to supercool the liquid and obtain
well-defined quasi-equilibrium properties for the metastable liquid
below the first order transition at $T_m$.
The energy and entropy curves for this liquid
exhibit fairly abrupt reproducible changes in slope at a rather
well-defined temperature, $k_B T_g\sim 0.4$ eV.
This suggests that the liquid phase
undergoes a glass transition at $k_B T_g \sim 0.4$ eV,
so $T_g \sim 0.75 T_m$.

It is interesting to note in Fig.\ 2
that extrapolations of  $E_a(T)$ and $S_a(T)$
from above $T_g$ would give negative
values at very low temperature.
This phenomenon is known as the Kauzmann paradox, and is
rather common for materials that form
structural glasses \cite{Kauz}.
Of course Fig.\ 2 shows that
the energy and entropy for $k_B T < 0.4  eV$
lie significantly above the curves extrapolated from higher $T$,
thereby avoiding a true,
unphysical paradox.

One can study this transition more quantitatively by calculating the
``time''-dependent structure factor \cite{Yip}
\begin{equation}
S(\vec{q},t)= \langle \rho(\vec{q},t')\rho(-\vec{q},t+t') \rangle  ~~,
\end{equation}
where $\rho(\vec{q},t')$ is the Fourier transform of the atomic
density at time $t'$,
$\rho (\vec{q},t')= \sum_{j=1}^{N}
\exp(i\vec{q} \cdot \vec{r}_j(t')) / \sqrt{N}$,
and the angle brackets denote
an average over $t'$.
In the liquid phase, $S(\vec{q} ,t)$
decays exponentially, $S(\vec{q},t)\sim \exp(-D|\vec{q}|^2 t)$, where $D$
is the diffusion constant of the liquid\cite{liq}.
In a glass phase, however, where there is frozen (albeit random)
structure, $S(\vec{q} ,t)$ in the thermodynamic limit
remains finite as $t \rightarrow \infty$,
decaying only because of the finite size of the simulated system.
We choose $|\vec{q}| = 2\pi/b_0 $, with $b_0$ the bond length of the crystal
structure, so that $S(\vec{q} ,t)$ will decay as rapidly as possible
in the liquid phase.  In addition to averaging over $t'$, we
average over many orientations
of $\vec{q}$ to get better statistics.  The results are
shown in Fig.\ 3, where the normalized time-dependent structure factor
$S(|\vec{q}|,t)/S(|\vec{q}|,0)$ is plotted versus time delay $t$,
for four different temperatures  around $T_g$.
(The ``time" units here are MC steps per atom.)

It can be seen from Fig.\ 3 that
the structure factor decays to zero for
$k_B T$=0.40 and 0.45  eV,
but remains finite for the lower temperatures, $k_B T$=0.30 and
0.35 eV.
Of course the number of accepted MC switching steps
decreases with decreasing temperature.
However, plotting the structure factor vs.\ {\it accepted} steps
gives a similar picture, leading to the same conclusion.
The simulations ran long enough to produce more than 2000 {\it accepted}
switching moves even for the lowest temperature,
$k_B T$=0.3 eV.

These results confirm the conclusion above,
that there is a well-defined glass transition
in the idealized CRN model \cite{Binder},
and this transition occurs at $k_B T_g\sim 0.40$ eV
for the specific network considered here.
Thus the CRN model has the following phase structure:
In equilibrium, as temperature decreases there is a first-order transition
from a disordered network ``liquid" phase to the crystalline phase.
The liquid phase may be supercooled below $T_m$, and at $T=T_g<T_m$
undergoes a transition into a metastable glassy phase.

It is interesting to note that the model glass phase
obtained by cooling the metastable liquid through $T_g$ has essentially
the same structure as was shown by WWW \cite{W3} to provide the best
available model for the structure of real a-Si.
Yet real a-Si is formed by radically different processes,
very far from equilibrium.
This suggests the existence of a rather well-defined
metastable amorphous phase, whose structure (after annealing)
is relatively independent of the kinetic history of the material.

As demonstrated in our calculation of $T_m$, the random network model
can easily be adapted to study the interface between
amorphous and crystalline phases.
To investigate the
effects of crystalline anisotropy, we prepared interfaces as before,
oriented parallel to the (100), (110), and (111) planes of the crystal.
We then equilibrated at $T=T_m$, the only
temperature for which the interface is stationary.
After initial transients,
the (100) and (110) interfaces develop \{111\} facets,
while the (111) interface remains flat.
For the (100) interface,
Fig.\ 4(a) shows the positions of all the Si atoms
at one instant of time.
It is clear that the interface is unstable --- the dotted lines
indicate two \{111\} facets which have formed.
Similar faceting occurs for the (110) interface.
Fig.\ 4(b) shows that, in contrast,
the (111) interface remains planar.
Whether such faceting occurs in real a-Si will depend on
the relative importance of thermodynamic and kinetic factors.

We can study the decay
of crystalline order near the interface in the amorphous phase
by calculating the local energy density, which is zero
for the perfect crystal and non-zero in the amorphous phase.
We find that the partial crystalline structure on the
amorphous side of the interface always
forms in double layers, consistent with the crystal structure
in the [111] direction.
This partial order
decays rapidly as one moves away from the interface,
with a decay length of 
roughly one (111) double layer.

In summary, we have studied the
properties of a random-network model
for amorphous materials.
While we have used parameters for a-Si,
the general behavior should be rather generic.
We demonstrated the existence of a glass transition
in an ideal network model.
We have also applied the
model to study the crystal-amorphous interface.
For Si the (111) interface was found to be stable,
but (110) and (001) interfaces are unstable against formation
of (111) facets.
This general approach should be applicable to a variety of systems
which maintain a nearly ideal network structure,
as well as for studies of generic aspects of network glasses.

D.V.\ acknowledges support of NSF Grant DMR-9613648.

\begin{figure}
\caption{Snapshots of projected atom positions
at time intervals of 200 MC steps per atom.
(a) $T=0.5$ eV$/k_B$ $(T<T_m)$, where the crystalline phase
eventually takes over the whole space.
(b) $ T=0.55$ eV$/k_B$ $(T>T_m)$, where the whole 
system becomes amorphized.}
\end{figure}

\begin{figure}
\caption{Energy, entropy, and free energy per atom versus temperature.
The break in slope
for energy and entropy at $T=T_g\sim 0.4$ eV$/k_B$ corresponds to a
glass transition.  ($T_0 \equiv 1$ $eV/k_B$
is introduced so that entropy can be plotted in units of energy.)}
\end{figure}

\begin{figure}
\caption{Normalized dynamical structure factor versus time
(in MC steps per atom)
for temperatures around
the glass transition.}
\end{figure}

\begin{figure}
\caption{Interface morphology for different orientations.  The
(100) interface develops (111) facets
(dotted lines). The atom positions are projected
onto the $(01\bar{1})$ plane.}
\end{figure}

\begin{references}
\bibitem{FcT} Since the phonon temperature is always zero,
deviations from $F_c(T)=0$ occur only because of
an exponentially small number of thermal lattice defects
in the crystal.
\end{references}
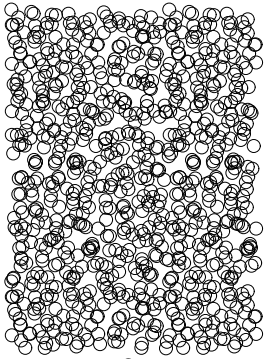
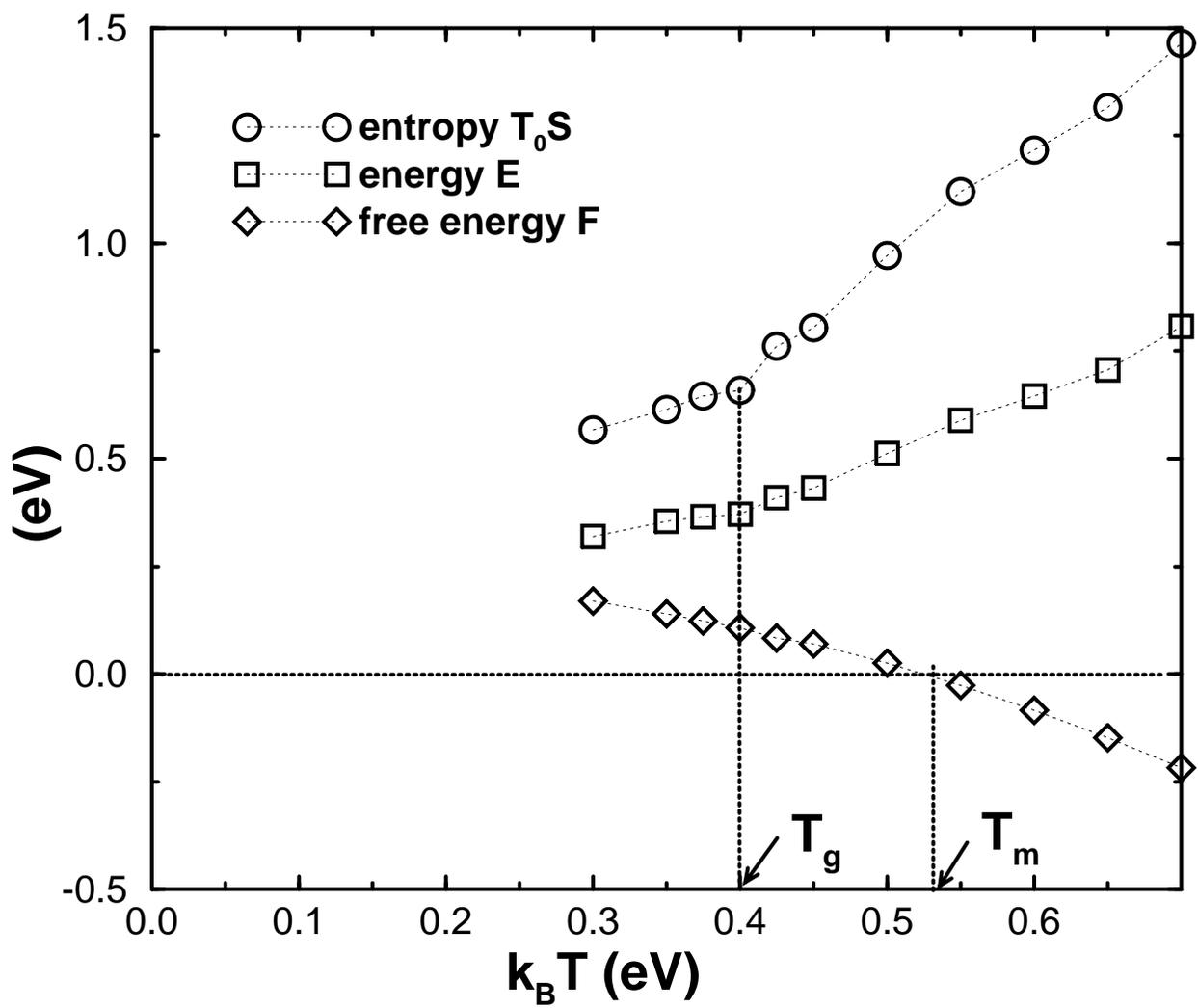
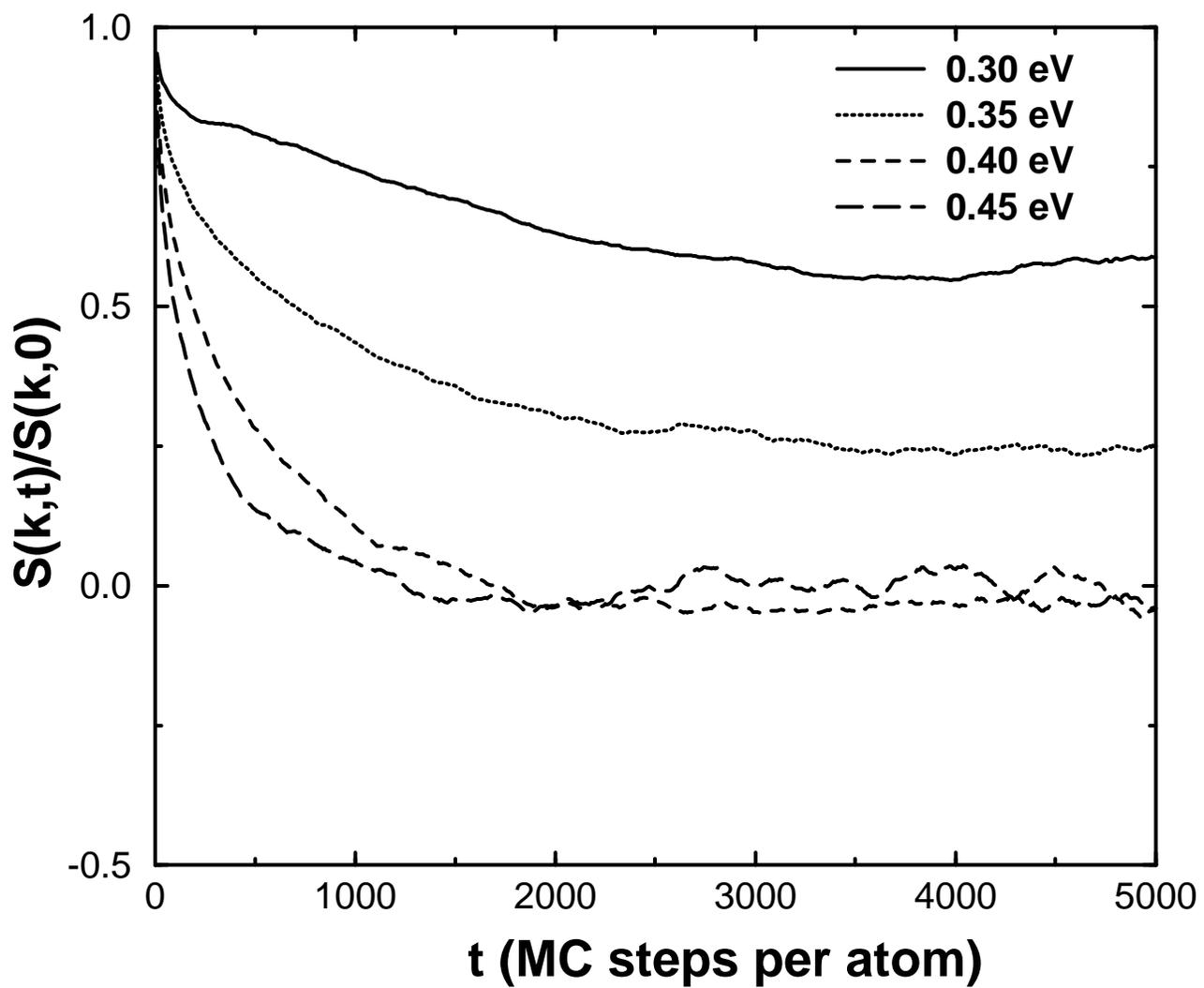
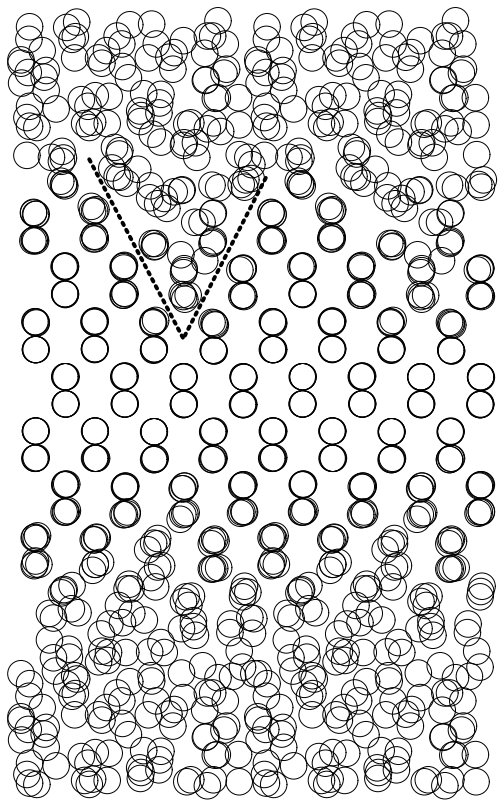
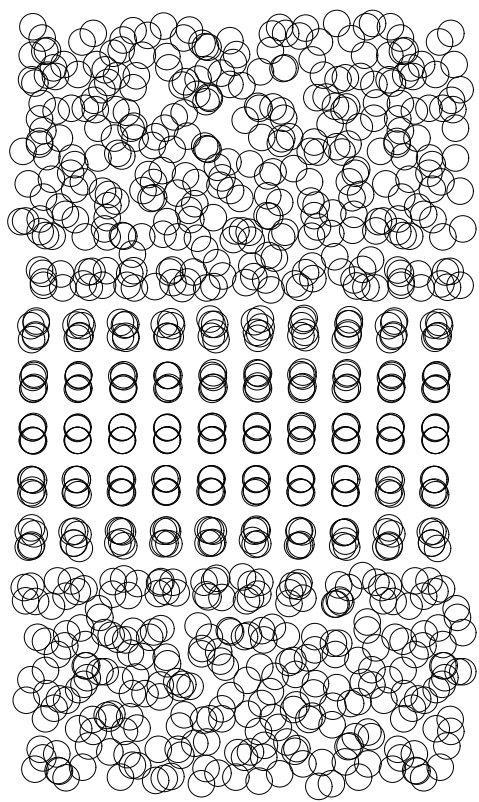
\end{document}